\documentstyle[11pt,epsfig,epsf,axodraw]{article}
\newcommand{\ba}{\begin{array}}
\newcommand{\ea}{\end{array}}
\newcommand{\bd}{\begin{displaymath}}
\newcommand{\ed}{\end{displaymath}}
\newcommand{\be}{\begin{equation}}
\newcommand{\ee}{\end{equation}}
\newcommand{\bea}{\begin{eqnarray}}
\newcommand{\eea}{\end{eqnarray}}

\def\q2 {q^2}

\begin{document}
\begin{center}

{\Large\bf  Signals of R-parity Violating Supersymmetry at a Muon Storage Ring}\\[15mm]

Anindya Datta\footnote{E-mail: anindya@mri.ernet.in}, 
Raj Gandhi\footnote{E-mail: raj@mri.ernet.in}, 
Biswarup Mukhopadhyaya\footnote{E-mail: biswarup@mri.ernet.in}\\
{\em Harish-Chandra Research Institute, Chhatnag Road, Jhusi, Allahabad - 211019.}\\[7mm]
* Poonam Mehta\footnote{E-mail: pmehta@physics.du.ac.in, 
delivered talk at XIV DAE Symposium on HEP, 18~-~22 December 2000.}\\
{\em Department of Physics and Astrophysics, University of Delhi, Delhi - 110007.}\\
\end{center}
\vskip 12pt
\begin{abstract}
  
          Neutrino oscillation signals at muon storage rings (MSR) can be 
  mimicked by supersymmetric (SUSY) interactions in an R-parity violating 
  scenario. We have investigated the $\tau$-appearance signals for both 
  long-baseline and near-site experiments, and concluded that the latter 
  is of great use in distinguishing between oscillation and  SUSY effects. 
  On the other hand, SUSY can cause a manifold increase in the event rate 
  for wrong-sign muons at a long-baseline setting, thereby providing us 
  with signatures of new physics.
  
\end{abstract}

\vskip 1 true cm
\setcounter{footnote}{0}
\def\baselinestretch{1.8}

The increasingly strong empirical indications of neutrino oscillations from
the observed solar and atmospheric neutrino deficits
have emphasised the need for their independent confirmation in accelerator and
reactor experiments. One of the actively discussed possibilities in this
connection is the proposal of a neutrino factory based on 
a MSR \cite{neufac} which can act as an intense source of collimated 
neutrinos impinging upon a fixed target. 
A $\mu^-$ ($\mu^+$) beam can thus produce both $\nu_\mu$ (${\bar \nu_\mu}$) 
and ${\bar \nu_e}$ ($\nu_e$), thereby providing an opportunity to test both
$\nu_e$-$\nu_\mu$ and $\nu_\mu$-$\nu_\tau$ oscillations which are the 
favoured solutions for the two anomalies mentioned above.
At a MSR with a $\mu^-$ beam, one expects a certain fraction 
of the $\nu_\mu$'s (${\bar \nu_e}$'s) to oscillate into 
$\nu_\tau$ (${\bar \nu_\mu}$), depending on the 
energy and the baseline length. Interaction of these $\nu_\tau$'s 
(${\bar \nu_\mu}$'s) with an {\it isonucleon} target will produce 
$\tau^-$($\mu^+$)-leptons, the detection of which may, in the simplest 
case, be interpreted as a signature of $\nu$-oscillation at a MSR.

However, the predicted rates of $\tau$-appearance or wrong-sign muons
in a given experimental setting can be significantly affected 
by non-standard interactions, i.e. to say that some non-oscillation 
physics can intervene and mimic the oscillation phenomena.
For example, it is possible for {\em unoscillated
$\nu_\mu$'s} to scatter into $\tau$'s in an R-parity violating
SUSY framework ({\em with $R~=~(-1)^{(3B + L + 2S)}$}), by
virtue of lepton-number violating trilinear couplings \cite{rparity}.
Also, such couplings can produce ${\bar \nu_\mu}$'s from
$\mu^{-}$-decay and thus give rise to $\mu^{+}$'s in the detector even
in the absence of oscillation. It is important to know the 
effects of such interactions for two reasons : {\it (i)} to look 
for {\em enhancement} in $\tau$ and wrong-sign muon event 
rates and thus to {\em uncover} new physics effects, and {\it (ii)} to see 
{\em to what extent} the signals supposedly coming from oscillation are 
faked by such new physics. In this paper, we have shown that one can 
answer both questions by combining long-baseline experiments with those 
in which one places the neutrino detectors at a short distance from 
the storage ring, where the oscillation probability gets suppressed 
by the baseline length.

Let us first consider $\tau$-appearance through oscillation.  For a
muon-neutrino with energy $E_\nu$ (in $GeV$) traversing a distance $L$
(in $km$), the probability of oscillation into a tau-neutrino is given by 
\begin{equation}
{\cal P}_{\nu_{\mu,e}\rightarrow \nu_\tau} = \sin^2 2\theta\,\, \sin^2\left[ 
1.27\, \Delta m^2\, {L\over E_\nu}\right]
\end{equation}
\noindent 
where $\Delta m^2$ is the mass-squared difference between the
corresponding physical states in $eV^2$, and $\theta$, the mixing
angle between flavours. 

In an R-parity violating SUSY scenario \cite{rparity} with broken 
lepton number, the trilinear terms in the superpotential  
(suppressing colour and $SU(2)$ indices) leads to interactions of the form 
\bea
{\cal {L}} &=&  \lambda'_{ijk} ~\big[ ~\tilde d^j _L \,\bar d ^k _R \nu^i _L
  + (\tilde d ^k_R)^\ast ( \bar \nu ^i_L)^c d^j _L +
   \tilde \nu ^i _L \bar d^k _R d ^j _L  \nonumber \\
& & ~~~~~~~ -\tilde e^i _L \bar d ^k _R u^j _L
- \tilde u^j _L \,\bar d ^k _R e^i _L 
-(\tilde d^k _R)^\ast (\bar e ^i _L)^c u^j _L \big] + h.c. \nonumber \\
& & + \lambda_{ijk} ~\big[ ~\tilde e^j _L \,\bar e ^k _R \nu^i _L
  + (\tilde e ^k_R)^\ast (\bar \nu ^i _L)^c e^j _L +
   \tilde \nu ^i _L \bar e^k _R e ^j _L  - (i \leftrightarrow j) \big] + h.c
\eea 
where, i,j and k are the generation indices .
At a neutrino factory, these interactions can affect the $\tau$ or wrong 
sign $\mu$ event rates in the following ways :

\begin{enumerate}
\item $\nu_\mu N \longrightarrow \tau^-  X$ via $\lambda'$-type interactions 
where ${\nu_\mu}$ is produced by the standard muon decay.

\item ${\bar \nu_\mu} N \longrightarrow \mu^+ X$ via standard charged current 
process where ${\bar \nu_\mu}$ is produced via $\lambda$-type interactions in the muon decay.

\item ${\nu_\tau} N \longrightarrow \tau^- X$via standard charged current 
process where ${\nu_\tau}$ is produced via $\lambda$-type interactions in the 
muon decay.

\item ${\bar \nu_e} N \longrightarrow \mu^+ X$via $\lambda'$-type interactions 
where ${\bar \nu_e}$ is produced by the standard muon decay.
\end{enumerate}
Here we present results for cases 1 and 2 above (see figure 1). The results 
for cases 3 and 4 are also qualitatively similar.

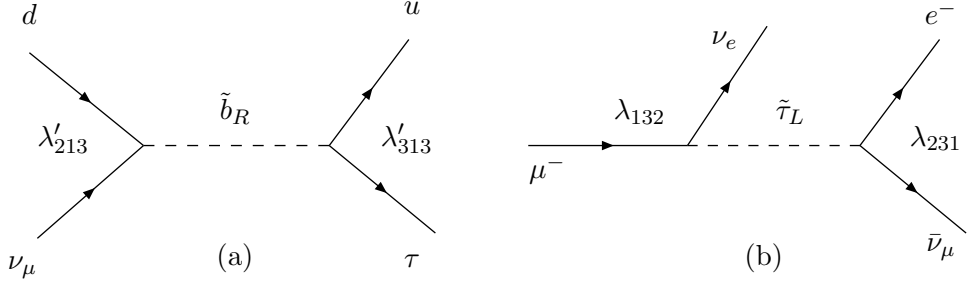
\begin{figure}
\hspace*{-.7in}

\begin{picture}(280,100)(0,0)
\vspace*{- 1in}
\ArrowLine(60,-5)(100,30)
\Text(60,-16)[r]{$\nu_\mu$}
\ArrowLine(57,65)(100,30)
\Text(60,80)[r]{$d$}
\DashLine(100,30)(170,30){4}
\Text(135,40)[b]{${\tilde b}_R$}
\ArrowLine(170,30)(200,70)
\Text(202,80)[b]{$u$}
\ArrowLine(170,30)(210,-3)
\Text(202,-16)[b]{$\tau$}
\Text(200,30)[b]{$\lambda '_{313}$}
\Text(70,30)[b]{$\lambda '_{213}$}
\Text(135,-15)[b]{(a)}

\ArrowLine(245,30)(305,30)
\Text(260,21)[r]{$\mu^-$}
\ArrowLine(305,30)(335,75)
\Text(325,70)[r]{$\nu_e$}
\DashLine(305,30)(370,30){4}
\Text(345,40)[b]{${\tilde \tau}_L$}
\ArrowLine(370,30)(400,70)
\Text(402,76)[b]{$e^-$}
\ArrowLine(370,30)(410,-3)
\Text(402,-10)[b]{${\bar\nu}_\mu$}
\Text(400,30)[b]{$\lambda _{231}$}
\Text(288,40)[b]{$\lambda _{132}$}
\Text(335,-15)[b]{(b)}
\end{picture}
\vskip .3in
\caption{ {\em Feynman diagrams for  processes producing (a) a $\tau$
or (b) a wrong sign muon in R-parity violating supersymmetry.}
}
\end{figure}
The standard model charged current cross-section for 
$\nu_\tau N \longrightarrow \tau^-  X$ is known \cite{sukanta}. 
Neglecting the left-right mixing in the squark sector , 
we get the following Fierz-transformed SUSY amplitudes 
for the most favourable quark-level processes ($\nu_\mu d \longrightarrow 
\tau^- u $ and $\nu_\mu {\bar u} \longrightarrow \tau^- {\bar d}$):
\bea
{\cal M}_{SUSY}(\nu_\mu \;d \longrightarrow \tau^- \; u) &=& 
\frac{\lambda'_{213}\lambda'_{313}}{2 (\hat s - m ^2_{\tilde b_R})}\;
\big[\bar u_{\tau} \gamma_{\mu} P_L u_{\nu_\mu}\big]\,
\big[\bar u_{u} \gamma^{\mu} P_L u_d \big] \nonumber \\
{\cal M}_{SUSY}(\nu_\mu \;\bar u \longrightarrow \tau^- \; \bar d) &=& 
\frac{\lambda'_{213}\lambda'_{313}}{2 (\hat t - m ^2_{\tilde b_R})}\;
\big[\bar u_{\tau} \gamma_{\mu} P_L u_{\nu_\mu}\big]\,
\big[\bar v_{u} \gamma^{\mu} P_L v_d \big] 
\eea
\noindent
where $m_{\tilde{b}}$ is the b-squark mass and the product of two 
$\lambda'$ couplings is taken to be real .
This tells us that the R-parity violating contributions can be included,
for $s << m^2_{\tilde{b}}$, on making the replacement 
$g^2/m_W^2 \rightarrow (g^2/m_W^2 + {(\lambda'_{213} \lambda'_{313})}/
{m_{\tilde b}^2)}$ in the standard expression for the charged current 
cross-section.
Although there are phenomenological bounds \cite{allanach} on the 
$L$-violating couplings, $\lambda'_{213}$ and $\lambda'_{313}$,  
no conclusive limit exists for the product ($\lambda'_{213}$$\lambda'_{313}$),
allowing it to be treated as a {\em free parameter when it comes 
to looking for experimental signals}.

From the plot of $\tau$-event rates versus the baseline length 
for a $50~GeV$ muon beam, we found that for baselines of 
length $\ge 200~km$, R-parity effects make a serious difference only 
when the couplings are close to their perturbative limits, while
for shorter baselines ($\simeq 100~km$), the R-parity violating effects 
are competitive even with values are on the order of the stand-alone bounds.
In figure 2(a) we've shown some plots of $\tau$-event rates ( using CTEQ4LQ 
parton distributions ) with a $1~kT$,
$2500~cm^2$ detector placed at a distance of $40~m$ from the storage ring. 
The standard model contribution to the $\tau$-production rates 
has been calculated assuming an oscillation probability
corresponding to $\Delta m_{23}^2~\simeq~5.0 \times 10^{-3}~eV^2$ and
$\sin^2 \theta_{23} ~=~1$, which is approximately the central region
of the solution space for the atmospheric $\nu_\mu$ deficit. 
$\lambda'$-type couplings are found to {\em enhance} the number of 
$\tau$-events to a level considerably higher than what the standard 
model predicts . The conclusion, therefore, is that {\em even when 
the couplings are well within the bounds for the stand-alone situation}, 
near-site effects arising from them lead to overwhelmingly large 
$\tau$-production, while for long-baseline experiments, 
contamination of the oscillation signals through R-violating 
interactions is appreciable when one goes beyond the limits for the 
stand-alone situation .
\begin{figure}[ht]
\centerline{ 
\epsfxsize=7.5cm\epsfysize=5.5cm\epsfbox{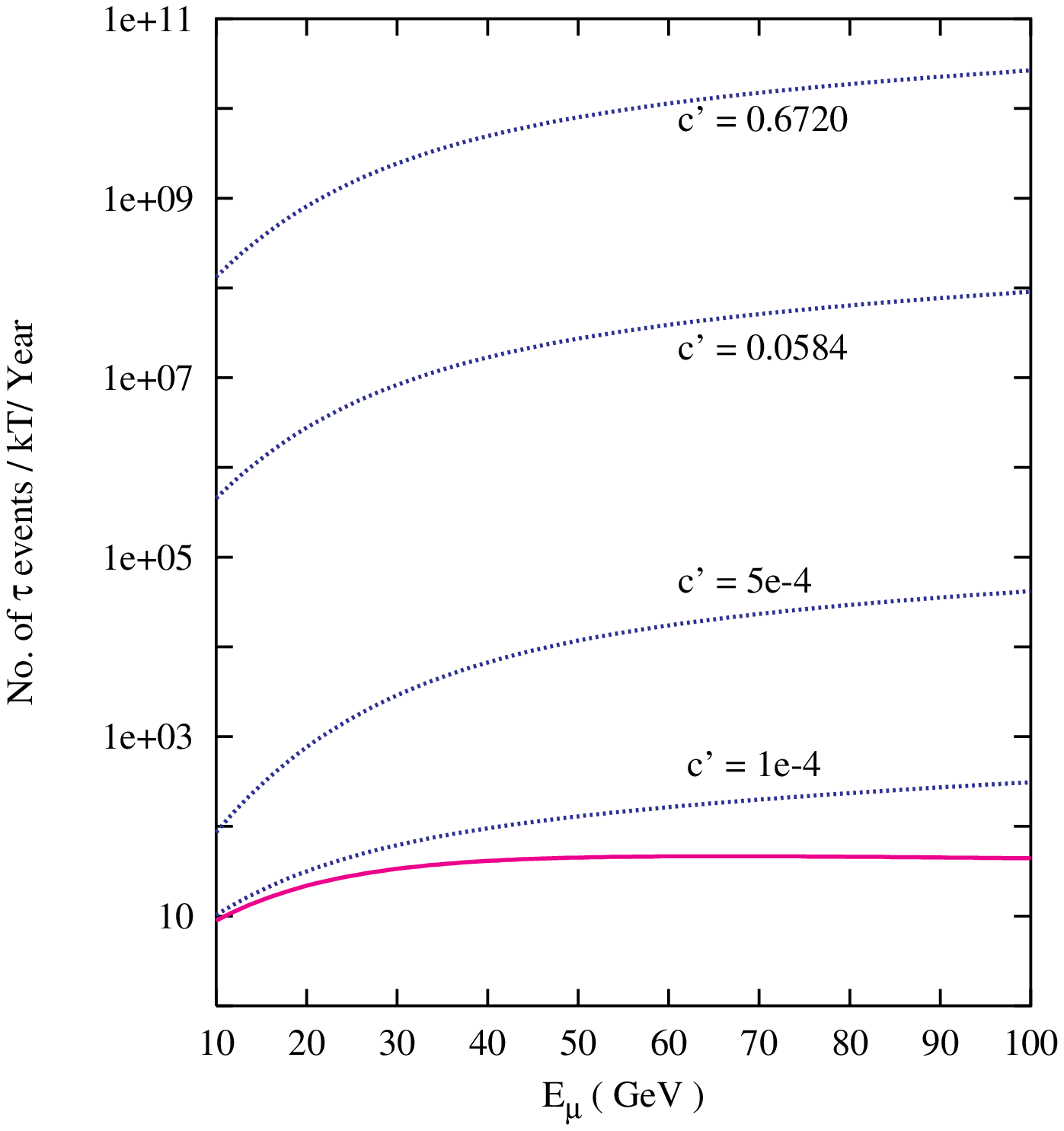}
        \hspace*{-1ex}
\epsfxsize=7.5cm\epsfysize=5.5cm
                     \epsfbox{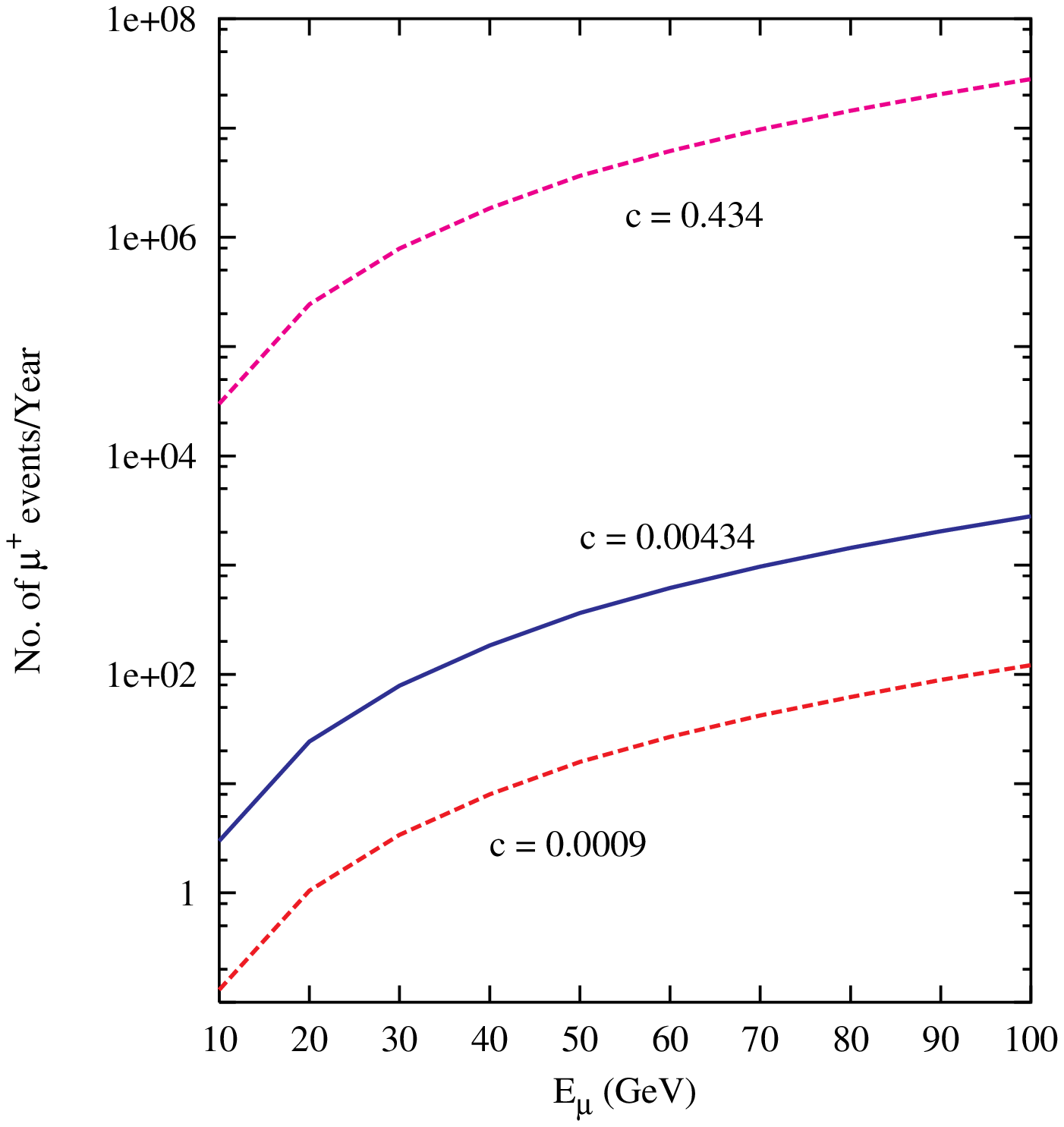}
}
\hskip 5cm
{\bf (a)}
\hskip 7cm
{\bf (b)}
\caption{(a){\em The $\tau$-event rate as function of the muon beam energy
for a near-site detector taking $N_\mu$ = $10^{20}$ per year and average 
$\tau$ detection efficiency = 30\%. The solid line shows the contribution 
from $\nu_\mu-\nu_\tau$ oscillation, using SK parameters (see text) ,
while the SUSY contributions are included to the dashed lines, using 
different values of $c' ~(\equiv \lambda'_{213} \lambda'_{313})$ and 
$m_{\tilde{b}} = 300~GeV$.} (b){\em The event rate for wrong-sign 
muons as a function of muon energy, for a baseline length of 250 km 
(K2K proposal) and a 10 kT detector of area 100 mt$^2$. Different values of 
$c ~(\equiv \lambda_{231} \lambda_{132})$ have been used, 
with $m_{\tilde \tau } = 100 ~GeV$.}}
\label{fig:fig2}
\end{figure}

Next, we consider wrong-sign muons produced due to R-parity violating
effects in {\em muon decays}, see figure 1(b).
The Mikhyev-Smirnov-Wolfenstein (MSW) solution to the solar neutrino 
problem with matter-enhanced $\nu_e$-$\nu_\mu$ oscillation requires a 
mass-splitting of $\simeq~10^{-5}~eV^2$ between the mass eigenstates.  
It has been found earlier that with a muon beam energy of upto 
$50~GeV$, and with standard charged current interactions, 
one can hardly expect to see any events given this kind
of mass-splitting, for any realistic baseline length. The situation is
even worse for the vacuum oscillation solution which requires $\Delta
m^2~\simeq~10^{-10}~eV^2$. Thus a {\em sizable} event rate for wrong-sign 
muons at a long-baseline experiment should be interpreted as a signal 
of some new effect, unless $\nu_e$-$\nu_\mu$ is {\it not} the solution 
of the solar neutrino puzzle.

In figure 2 (b) we've plotted the event rates for a typical ICANOE-type 
detector as functions of the muon beam energy, for a baseline of length  
 $250 kms$ using different values of the product 
$\lambda_{231} \lambda_{132}$, 
including those considerably smaller than the most stringent 
phenomenological limits. 
Even with conservative choices of the interaction strengths, a 
clear prediction of ten to several hundred events can be observed for 
$E_\mu~\simeq~50~GeV$ in the SUSY case, while no events are expected 
so long as masses and mixing in the $\nu_\mu$-$\nu_e$ sector offer 
a solution to the solar deficit.

In conclusion, we have investigated the effects of the R-parity
violating trilinear couplings on signals of $\nu$-oscillations 
at a MSR. We have found that, while new couplings have to be on the 
higher side to show a detectable enhancement in the $\tau$-appearance 
rate with long baselines, even tiny R-violating couplings can lead to very 
large no. of $\tau$'s at a near-site detector setting, much in excess 
to that expected via oscillation. Near-site experiments can thus be 
recommended for isolating the new physics effects that {\em mimic} 
the signals of $\nu$-oscillation . On the other hand, a class of 
R-violating interactions, with strengths well below their current 
experimental limits, can be responsible for an enhanced rate of 
$\mu^+$ even at a long-baseline experiment. Since the solution space for 
the solar $\nu$-puzzle does not permit such event rates, such muons, 
if observed at a MSR, can therefore be greeted as harbingers of some 
new physics.

\noindent
{\bf Acknowledgement:}P.M. acknowledges financial support from the Council 
for Scientific and Industrial Research, India.

\newcommand{\plb}[3]{{Phys. Lett.} {\bf B#1,} #2 (#3)}                  %
\newcommand{\prl}[3]{Phys. Rev. Lett. {\bf #1,} #2 (#3)}        %
\newcommand{\rmp}[3]{Rev. Mod.  Phys. {\bf #1,} #2 (#3)}             %
\newcommand{\prep}[3]{Phys. Rep. {\bf #1,} #2 (#3)}                     %
\newcommand{\rpp}[3]{Rep. Prog. Phys. {\bf #1,} #2 (#3)}             %
\newcommand{\prd}[3]{Phys. Rev. {\bf D#1,} #2 (#3)}                    %
\newcommand{\prc}[3]{{Phys. Rev.}{\bf C#1,} #2 (#3)}  
\newcommand{\np}[3]{Nucl. Phys. {\bf B#1,} #2 (#3)}                    %
\newcommand{\npbps}[3]{Nucl. Phys. B (Proc. Suppl.) 
           {\bf #1,} #2 (#3)}                                           %
\newcommand{\epj}[3]{Euro. Phys. J{\bf C#1,} #2 (#3)}                  %

\newcommand{\sci}[3]{Science {\bf #1,} (#3) #2}                 %
\newcommand{\zp}[3]{Z.~Phys. C{\bf#1,} #2 (#3)}                 %
\newcommand{\mpla}[3]{Mod. Phys. Lett. {\bf A#1,} #2 (#3)}             %
 \newcommand{\apj}[3]{ Astrophys. J.\/ {\bf #1,} #2 (#3)}       %
\newcommand{\astropp}[3]{Astropart. Phys. {\bf #1,} #2 (#3)}            %
\newcommand{\ib}[3]{{ ibid.\/} {\bf #1,} #2 (#3)}                    %
\newcommand{\nat}[3]{Nature (London) {\bf #1,} (#3) #2}         %
 \newcommand{\app}[3]{{ Acta Phys. Polon.   B\/}{\bf #1,} (#3) #2}%
\newcommand{\nuovocim}[3]{Nuovo Cim. {\bf C#1,} (#3) #2}         %
\newcommand{\yadfiz}[4]{Yad. Fiz. {\bf #1,} (#3) #2;             %
Sov. J. Nucl.  Phys. {\bf #1,} #3 (#4)]}               %
\newcommand{\jetp}[6]{{Zh. Eksp. Teor. Fiz.\/} {\bf #1,} (#3) #2;
           {JETP } {\bf #4} (#6) #5}%
\newcommand{\philt}[3]{Phil. Trans. Roy. Soc. London A {\bf #1} #2  
        (#3)}                                                          %
\newcommand{\hepph}[1]{(hep--ph/#1)}           %
\newcommand{\hepex}[1]{ (hep--ex/#1)}           %
\newcommand{\astro}[1]{(astro--ph/#1)}         %

\end{document}